# Cultural route to the emergence of linguistic categories

Andrea Puglisi (1), Andrea Baronchelli (1,2) & Vittorio Loreto (1)

*(1)"La Sapienza" University, Physics Department, P.le A. Moro 2, 00185 Rome ITALY*

*(2) Departament de Física i Enginyeria Nuclear, Universitat Politècnica de Catalunya, Campus Nord, Mòdul B4, 08034 Barcelona, Spain*

Andrea.Puglisi@roma.infn.it, Andrea.Baronchelli@roma1.infn.it, Vittorio.Loreto@roma1.infn.it

**Abstract**

Categories provide a coarse grained description of the world. A fundamental question is whether categories simply mirror an underlying structure of nature, or instead come from the complex interactions of human beings among themselves and with the environment. Here we address this question by modelling a population of individuals who co-evolve their own system of symbols and meanings by playing elementary language games. The central result is the emergence of a hierarchical category structure made of two distinct levels: a basic layer, responsible for fine discrimination of the environment, and a shared linguistic layer that groups together perceptions to guarantee communicative success. Remarkably, the number of linguistic categories turns out to be finite and small, as observed in natural languages.

Categories are fundamental to recognize, differentiate and understand the environment. According to Aristotle, categories are entities characterized by a set of properties which are shared by their members [1]. A recent wave in cognitive science, on the other hand, has operated a shift in viewpoint from the object of categorization to the categorizing subjects [2,3]: categories are culture-dependent conventions shared by a given group. In this perspective, a crucial question is how they come to be accepted at a global level without any central coordination [4,5,6-8]. The answer has to be found in communication, that is the ground on which culture exerts its pressure. An established breakthrough in language evolution [4,9-11] is the appearance of linguistic categories, i.e. a shared repertoire of form-meaning associations in a given environment [2,3,5,12-15]. Different individuals may in principle perceive, and even conceptualize, the world in very different ways, but they need to align their linguistic ontologies in order to understand each other.

In the past there have been many computational and mathematical studies addressing the learning procedures for form-meaning associations [16-17]. From the point of view of methodology, the evolutionary scheme, based on the maximization of some fitness functions, has been extensively applied [18,19]. Recent years, however, have shown that also the orthogonal approach of self-organisation can be fruitfully exploited in multi-agent models for the emergence of language [6-8]. In this context, a community of language users is viewed as a complex dynamical system which has to develop a shared communication system [20,21]. In this debate, a still open problem concerns the emergence of a small number of forms out of a diverging number of meanings. For example the few "basic colour terms", present in natural languages, coarse-grain an almost infinite number of perceivable different colours [22-24].

Following this recent line of research, our work shows that an assembly of individuals with basic communication rules and without any external supervision, may evolve an initially empty set of categories, achieving a non-trivial communication system characterized by a few linguistic categories. To probe the hypothesis that cultural exchange is sufficient to this extent, individuals in our model are never replaced (as in evolutionary schemes [18,19]), the only evolution occurring in their internal form-meaning association tables, i.e. their "mind". The individuals play elementary language games [25,26] whose rules, hard-wired in individuals, constitute the only knowledge initially shared by the population. They are also capable of perceiving analogical stimuli and communicating with each others [6,7].

Our model involves a population of $N$ individuals (or players), committed in the categorization of a single analogical perceptual channel, each stimulus being represented as a real-valued number ranging in the interval $[0,1]$. Here we identify categorization as a partition of the interval $[0,1]$ in discrete sub-intervals, from now onwards denoted as perceptual categories [27]. Each individual has a dynamical inventory of form-meaning associations linking perceptual categories (meanings) to words (forms), representing their linguistic counterpart. Perceptual categories and words associated to them co-evolve dynamically through a sequence of elementary communication interactions, simply referred as games. All players are initialised with only the trivial perceptual category $[0,1]$, with no name associated to it. At each time step a pair of individuals (one playing as speaker and the other as hearer) is selected and presented with a new ``scene'', i.e. a set of $M \geq 2$ objects (stimuli), denoted as $o_i \in [0,1]$ with $i \in [1, M]$. The speaker discriminates the scene and names one object and the hearer tries to guess the named object. A correct guess makes the game successful. Based on game's outcomes both individuals update their category boundaries and the inventory of the associated words. A detailed description of the game is given in Fig. 1.

The perceptive resolution power of the individuals limits their ability to distinguish objects/stimuli that are too close to each other in the perceptual space: in order to take this into account, we define a threshold $d_{min}$, inversely proportional to their resolution power. In a given scene the $M$ stimuli are chosen to be at a distance larger than this threshold, i.e. $|o_i - o_j| > d_{min}$ for every pair $(i, j)$. Nevertheless, objects presented in different games may be closer than $d_{min}$. The way stimuli are randomly chosen characterizes the kind of simulated environment: simulations will be presented both with a homogeneous environment (uniform distribution in $[0,1]$) and more natural environments (e.g., without loss of generality [27], the distributions of the hue sampled from pictures portraying natural landscapes).

A resume of the main results of our experiments is given in Fig. 2. The evolution of the population presents two main stages: 1) a phase where players do not understand each other, followed by 2) a phase where communication has reached full success thanks to the emergence of a common language, still with evolving perceptual categories. The first phase is marked by the growth and decline of synonymy, see Fig. 2a. Synonymy, in the context of the ``naming game'' (an individual object to be named), has been already studied [8], and a similar evolution was observed and explained. All individuals, when necessary, create new words with zero probability of repetition: this leads to an initial growth of the vocabulary associated to each perceptual category. New words are spread through the population in later games and, whenever a word is understood by both players, other competing words for the same category are forgotten. This eventually leads to only one word per category. During the growth of the dictionary the success rate, see Fig. 2b, is very small. The subsequent reduction of the dictionary corresponds to a growing success rate which reaches its maximum value after synonymy has disappeared. In all our numerical experiments the final success rate overcomes 80% and in most of them goes above 90%, weakly increasing with the final number of perceptual categories. Success is reached in a number of games per player of the order of $5 \cdot 10^2$, logarithmically depending on $N$, and it remains constant hereafter.

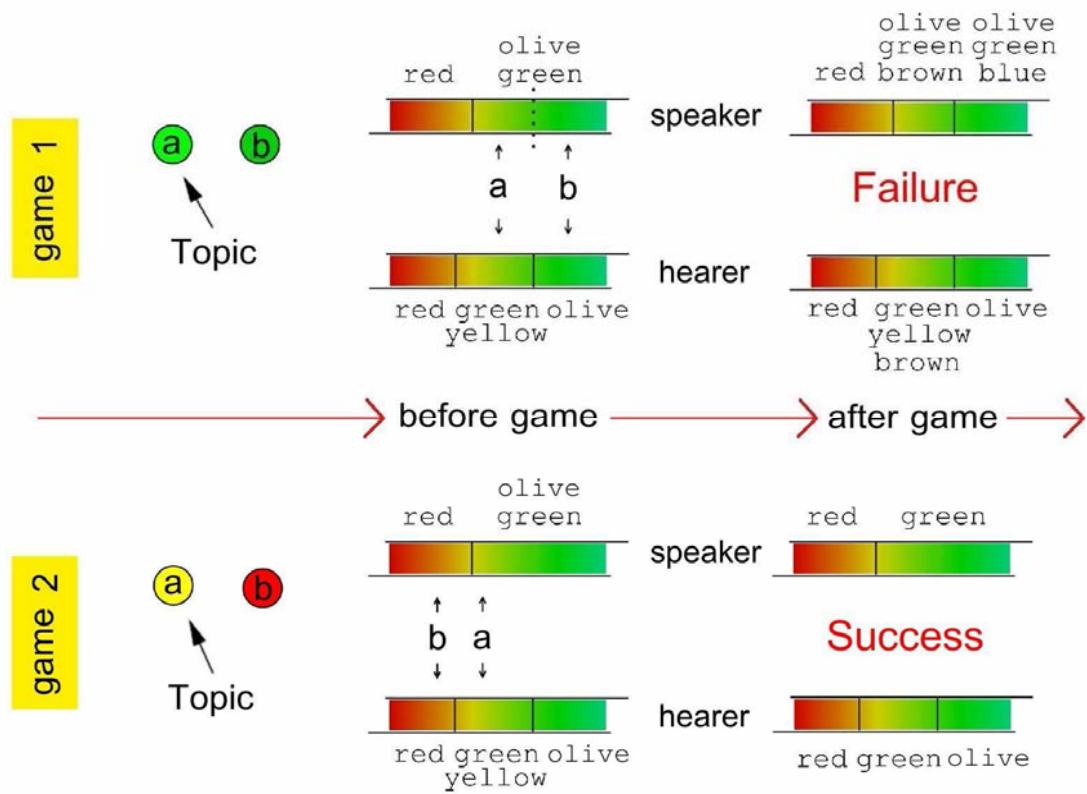

**Fig.1**: **Rules of the game.** A pair of examples representing a failure (game 1) and a success (game 2), respectively. In a game, two players are randomly selected from the population. Two objects are presented to both players. The speaker selects the topic. In **game 1** the speaker has to discriminate the chosen topic ("a" in this case) by creating a new boundary in her rightmost perceptual category at the position $(a+b)/2$. The two new categories inherit the words-inventory of the parent perceptual category (here the words "green" and "olive") along with a different brand new word each ("brown" and "blue"). Then the speaker browses the list of words associated to the perceptual category containing the topic. There are two possibilities: if a previous successful communication has occurred with this category, the last winning word is chosen; otherwise the last created word is selected. In the present example the speaker chooses the word "brown", and transmits it to the hearer. The outcome of the game is a failure since the hearer does not have the word "brown" in her inventory. The speaker unveils the topic, in a non-linguistic way (e.g. pointing at it), and the hearer adds the new word to the word inventory of the corresponding category. In **game 2** the speaker chooses the topic "a", finds the topic already discriminated and verbalizes it using the word "green" (which, for example, may be the winning word in the last successful communication concerning that category). The hearer knows this word and therefore points correctly to the topic. This is a successful game: both the speaker and the hearer eliminate all competing words for the perceptual category containing the topic, leaving "green" only. In general when ambiguities are present (e.g. the hearer finds the verbalized word associated to more than one category containing an object), these are solved making an unbiased random choice.

The set of perceptual categories of each individual follows a somewhat different evolution (see dashed lines in Fig. 2c). The first step of each game is, in fact, the discrimination stage where the speaker (possibly followed by the hearer) may refine her category inventory in order to distinguish the topic from the other objects. The growth of the number of perceptual categories $n_{perc}$ of each individual is limited by the resolution power: in a game two objects cannot appear at a distance smaller than $d_{min}$ and therefore $n_{perc} < 2/d_{min}$. The minimal distance also imposes a minimum number of categories $1/d_{min}$ that an individual must create before her discrimination process may stop. The average number of perceptual categories per individual, having passed $1/d_{min}$, grows sub-logarithmically and for many practical purposes it can be considered constant.

The success rate is expected to depend on the alignment of the category inventory among different individuals. The degree of alignment of category boundaries is measured by an *overlap function O* (defined in the Appendix) which returns a value proportional to the degree of alignment of the two category inventories, reaching its maximum unitary value when they exactly coincide. Its study, see dashed curves of

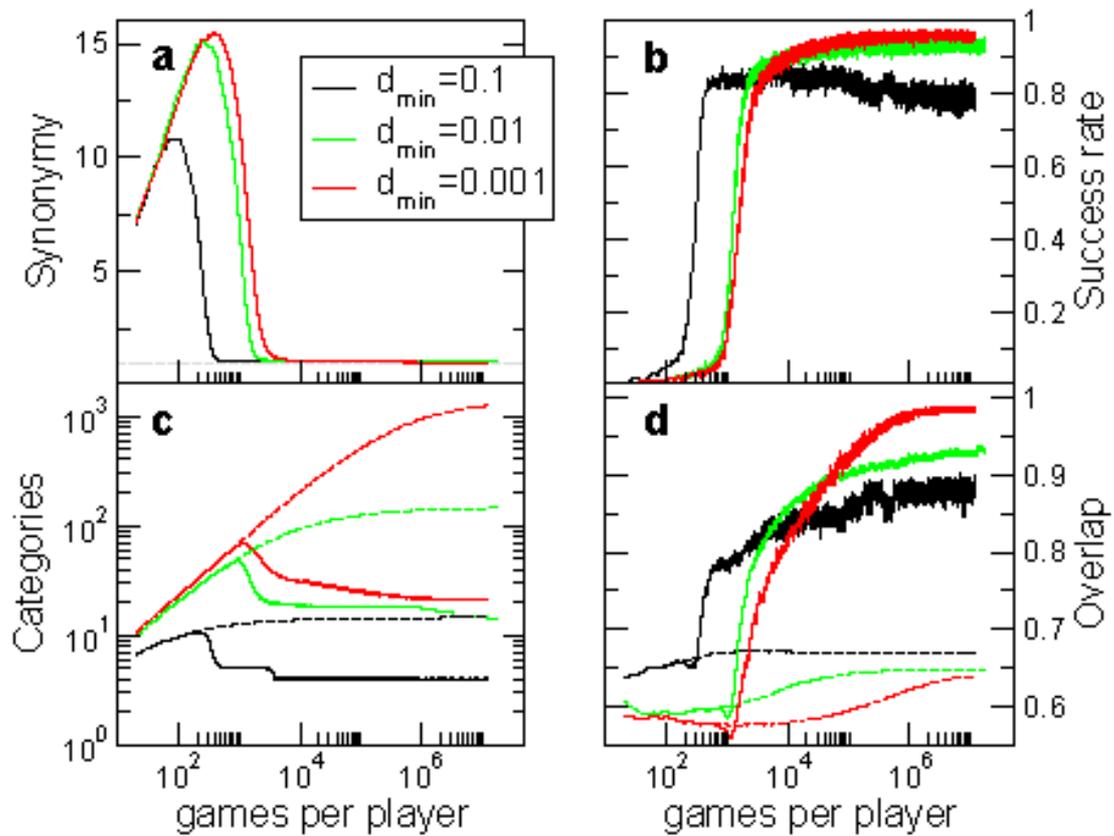

**Fig.2: Results of the simulations with** $N = 100$ **and different values of** $d_{min}$:
**a)** Synonymy, i.e. average number of words per category; **b)** Success rate measured as the fraction of successful games in a sliding time windows 1000 games long; **c)** Average number of perceptual (dashed lines) and linguistic (solid lines) categories per individual; **d)** Averaged overlap, i.e. alignment among players, for perceptual (dashed curves) and linguistic (solid curves) categories.

Fig. 2d, shows that alignment grows with time and saturates to a value which is, typically, in between 60%-70%, i.e. quite smaller than the communicative success. This observation immediately poses a question: given such a strong misalignment among individuals, why is communication so effective?

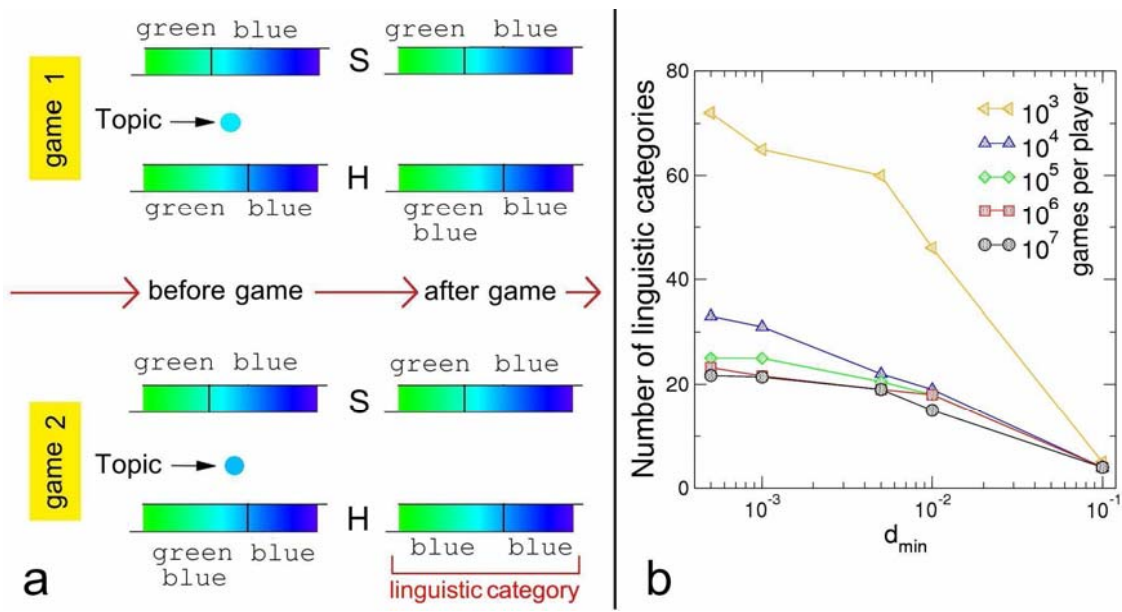

**Fig.3**: **Saturation in the number of linguistic categories**: **a)** A "word contagion" phenomenon occurs whenever the topic falls in a gap between two misaligned categories of two playing individuals. In the shown examples two individuals play two successive games. In game 1 the speaker (S) says "blue" and the hearer (H), unable to understand, adds "blue" as a possible word for her leftmost category; successively (game 2) the speaker repeats "blue" and the hearer learns this word as the definitive name for that perceptual category; both left and right perceptual categories of the hearer are now identified by the same name "blue" and they can be considered (for the purpose of communication) as a single linguistic category; **b)** Final number of linguistic categories as a function of $d_{min}$ at different times, with $N=100$. As the time increases the number of linguistic categories saturates. At large times, for small $d_{min}$, the number of linguistic categories becomes independent of $d_{min}$ itself. Concerning size dependence, only a weak (logarithmic) dependence on N, not shown, is observed.

The answer has to be found in the analysis of polysemy, i.e. the existence of perceptual categories identified by the same unique word. Misalignment, in fact, induces a ``word contagion'' phenomenon. With a small but non zero probability, two individuals with similar, but not exactly equal, category boundaries, may play a game with a topic falling in a misalignment gap, as represented in Fig. 3a. In this way a word is copied to an adjacent perceptual category and, through a second occurrence of a similar event, may become the unique name of that category. Interfering events may occur in between: it is always possible, in fact, that a game is played with a topic object falling in the bulk of the category, where both players agree on its old name, therefore cancelling the contagion. With respect to this cancelling probability, some gaps are too small and act as almost perfectly aligned boundaries, drastically reducing the probability of any further contagion. Thus, polysemy needs a two-step process to emerge, and a global self-organised agreement to become stable. On the other hand, polysemy guarantees communicative success: perceptual categories that are not perfectly aligned tend to have the same name, forming true linguistic categories, much better aligned among different individuals. The topmost curve of Fig. 2d, displays the overlap function measured considering only boundaries between categories with different names: it is shown to reach a much higher value, even larger than 90%.

The appearance of linguistic categories is the evidence of a coordination of the population on a higher hierarchical level: a superior linguistic structure on top of the individual-dependent, finer, discrimination layer. The linguistic level emerges as totally self-organised and is the product of the (cultural) negotiation process among the individuals. The average number of linguistic categories per individual, $n_{ling}$, Fig. 2c (solid curves), grows together with $n_{cat}$ during the first stage (where communicative success is still lacking), then decreases and stabilises to a much lower value. Some configurations of both category layers, at a time such that the success rate has overcome 95%, are presented in Fig. 4, using different sets of external stimuli.

The analysis, resumed in Fig.3b, of the dependence of $n_{ling}$ on $d_{min}$ for different times, makes our findings robust and, to our knowledge, unprecedented. As the resolution power is increased, i.e. as $d_{min}$ is diminished, the asymptotic number of linguistic categories becomes less and less dependent upon $d_{min}$ itself. Most importantly, even if any state with $n_{ling} > 1$ is not stable, we have the clear evidence of a saturation with time, in close resemblance with metastability in glassy systems [28,29]. This observation allows to give a solution to the long-standing problem of explaining the finite (and small) number of linguistic categories $n_{ling}$. In previous pioneering approaches [6,7] the number of linguistic categories $n_{ling}$ was trivially constrained (with a small range of variability) by $d_{min}$, with a relation of the kind $n_{ling} \propto 1/d_{min}$, implying a divergence of $n_{ling}$ with the resolution power. In our model we have a clear indication of a finite $n_{ling}$ even in the continuum limit, i.e. $d_{min} \to 0$, corresponding to an infinite resolution power.

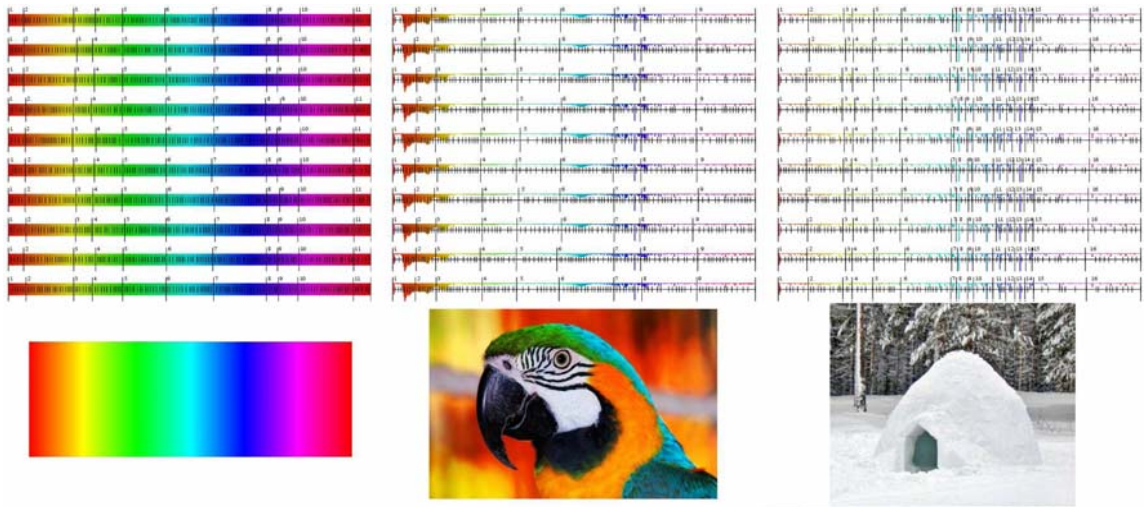

**Fig.4**: **Categories and the pressure of environment.** Inventories of 10 individuals randomly picked up in a population of $N = 100$ players, with $d_{min} = 0.01$, after $10^7$ games. For each player the configuration of perceptual (small vertical lines) and linguistic (long vertical lines) category boundaries is superimposed to a coloured histogram indicating the relative frequency of stimuli. The labels indicate the unique word associated to all perceptual categories forming each linguistic category. Three cases are presented: one with uniformly distributed stimuli (**a**) and two with stimuli randomly extracted from the hue distribution of natural pictures (**b and c**). One can appreciate the perfect agreement of category names, as well as the good alignment of linguistic category boundaries. Moreover, linguistic categories tend to be more refined in regions where stimuli are more frequent: an example of how the environment may influence the categorization process.

With the help of an extensive and systematic series of simulations we have shown that a simple negotiation scheme, based on memory and feedback, is sufficient to guarantee the emergence of a self-organised communication system which is able to discriminate objects in the world, requiring only a small set of words. Individuals alone are endowed with the ability of forming perceptual categories, while cultural interaction among them is responsible for the emergence and alignment of linguistic categories. Our model reproduces a typical feature of natural languages: despite a very high resolution power, the number of linguistic categories is very small. For instance, in many human languages, the number of "basic colour terms" used to categorize colours usually amounts to about ten [22-24], while the light spectrum resolution power of our eyes is evidently much higher. Note that in our simulations we observe a *reduction*, with time, of the number of linguistic categories toward the final plateau. The experimental evidence [30], collected in empirical studies on colour categorization, of a *growth* of the number of categories from simpler to more complex societies could be, in our opinion, an effect of the increased number $N$ of players involved in the communicative process. Finally we believe that these results could be important both from the point of view of language evolution theories, possibly leading to a quantitative comparison with real data (e.g. different populations sizes and ages), and from the point view of applications (e.g. emergence of new communication systems in biological, social and technological contexts [31,32]).

**Appendix**

The degree of alignment of category boundaries is measured by the following ``overlap'' $O$ function:

$$O = 2\sum_{\{i<j\}} \frac{o_{ij}}{N(N-1)} \text{ with } o_{ij} = \frac{2\sum_{c_i^j}(1-lc_i^i)^2}{\sum_{c_i}(1-lc_i)^2 + \sum_{c_i}(1-lc_j)^2},$$

where $lc$ is the width of category $c$, $c_{ii}$ is one of the categories of the $i$-th player, and $c_{ij}$ is the generic category of the category ``intersection'' set obtained considering all the boundaries of both players $i$ and $j$. The function $o_{ij}$ returns a value proportional to the degree of alignment of the two category inventories, reaching its maximum unitary value when they exactly coincide.

[27]  This approach can also be extended to categories with prototypes and fuzzy boundaries, for instance adding a weight structure upon it. Typical proposals in literature, such as prototypes with a weight function equal to the inverse of the distance

from the prototype [7], are exactly equivalent to our ``rigid boundaries'' categories. Moreover, all the results of our experiment can be easily generalized to multi-dimensional perceptual channels, provided an appropriate definition of category domains is given. It should be kept in mind that the goal of our work is to investigate why the continuum of perceivable meanings in the world is organized, in language, in a finite and small number of subsets with different names, with a no immediate (objective) cause for a given partition with respect to other infinite possibilities. Apart from the evident example of the partition of the continuous light spectrum in a small number of "basic color terms", this phenomenon is widespread in language: one can ask, for example, what objective differences allow to distinguish a *cup* from a *glass*; one can present a multi-dimensional continuum of objects able to "contain a liquid" (including also objects given as a prize), but a natural discontinuity between *cups* and *glasses* does not appear; our model, even reducing the phenomenon to the case of a 1-dimensional continuum, unveils a mechanism which can be easily extended to any kind of space, once it has been provided with a topology. The mechanism we propose for the discrete partition in linguistic subsets (categories) does not depend on the exact nature of this topology, which is of course a fundamental, yet different, matter of investigation.

**Acknowledgements:** The authors wish to thank C. Cattuto, G. Andrighetto, A. Baldassarri, E. Polizzi di Sorrentino, L. Steels, T. Belpaeme, J. De Beule and B. De Vylder for many interesting discussions and suggestions. This research has been partly supported by the EU under RD contract IST-1940 (ECAgents).

**Competing Interests statement**: The authors declare that they have no competing financial interests.